\documentclass{ws-ijmpd}
\usepackage[super,compress]{cite}
\usepackage{bm,amsmath,color,hyperref}

\begin{document}

\markboth{Arun and Pai}
{Tests of General Relativity and Alternative theories of gravity using GW observations}

\catchline{}{}{}{}{}
%
%%%%%%%%%%%%%%%%%%%%%%%%%%%%%%%%%%%%%%%%%%%%%%%%%%%%%%%%%%%%%%%%%%%%

\title{TESTS OF GENERAL RELATIVITY AND ALTERNATIVE THEORIES OF GRAVITY USING GRAVITATIONAL WAVE OBSERVATIONS}

%\author{FIRST AUTHOR\footnote{Typeset names in
%8~pt roman, uppercase. Use the footnote to indicate the
%present or permanent address of the author.}}
\author{K G ARUN}
\address{Chennai Mathematical Institute\\
Siruseri, Tamilnadu, 603103, India\\kgarun@cmi.ac.in}

\author{ARCHANA PAI}

\address{IISER Thiruvananthapuram, \\Computer Science
Building, College of Engineering Trivandrum Campus,\\ Trivandrum-695016
Kerala, India\\archana@iisertvm.ac.in}

\maketitle

%\begin{history}
%\received{Day Month Year}
%\revised{Day Month Year}
%\end{history}

\begin{abstract}
Gravitational Wave (GW) observations of coalescing compact binaries will be unique probes of strong-field, dynamical aspects of relativistic gravity.
We present a short review of various schemes proposed in the literature to test General Relativity (GR) and
alternative theories of gravity using inspiral waveforms. 
Broadly these schemes may be classified into two types: model dependent and model independent.
In the model dependent category, GW observations are compared against a specific waveform model representative of a particular theory
or a class of theories like Scalar-Tensor theories, Dynamical Chern-Simons theory and Massive graviton theories. Model independent
tests are attempts to write down a parametrised gravitational waveform where the free parameters take different values for different
theories and (at least some of) which can be constrained by GW observations. We revisit some of the proposed bounds in the case of downscaled LISA configuration (eLISA) and compare them with the  original LISA configuration.  We also compare the expected bounds on alternative theories of gravity from ground-based and space-based detectors and find that space-based GW detectors can test GR and other theories of gravity with unprecedented accuracies. We then focus on a recent proposal to use singular value decomposition (SVD) of the Fisher information matrix to improve the accuracies with which Post-Newtonian (PN) theory can be tested. We extend those results to the case of space based detector eLISA and discuss its implications. 
\end{abstract}


\ccode{PACS numbers: 04.25.Nx, 95.55.Ym, 04.30.Db, 97.60.Lf, 04.30.-w, 04.80.Cc and 04.80.Nn}

\tableofcontents

\section{Introduction}
General Relativity (GR) is the most studied and well-tested theory of gravity. It has been tested to extremely good accuracies 
by various solar system tests (weak-field tests) and binary pulsar observations (strong-field tests)(see Ref.\cite{Wthexp,Will05LivRev,Stairs03} for reviews). Despite these successes,
there is motivation to test the theory in regimes which are not accessible at present. Firstly, there is no
reason why GR is the fundamental theory of gravity at all scales. Though solar system and binary pulsar observations have tested GR to very good 
accuracies, there can be deviations from GR in the very strong field regime which may not be seen in the solar system or binary pulsar observations. There are unresolved problems of singularities in the theory and a complete quantum theory of gravity still remains to be a distant goal.
 Finally, even if it is the correct description of gravity, one would like to quantify the statement.
%%%%%%%%%%%%%%%%%%%%%%%%%%%%%%%%% 
\subsection{Tests of GR and Alternative theories of gravity using GW observations}
%%%%%%%%%%%%%%%%%%%
Gravitational Waves from compact binaries consisting of neutrons stars (NS) and/or black holes (BH) are unique probes of
strong field aspects of relativistic theories of gravity (See Ref.\cite{SathyaSchutzLivRev09} for a review of the field of GW astronomy. See Ref.~\cite{KambleKaplanRev12} for an overview of electromagnetic counterparts associated with compact binaries). The late stages of the evolution of the binary, from which observable
GWs are expected, are highly relativistic. The velocity of the binary, during the late stages, can be very close to the speed of light as opposed to binary pulsars which have speeds $\sim 10^{-3}c$. Hence GWs from inspiralling compact binaries directly carry the signatures of highly nonlinear, strong-field dynamics of the space time. Hence it is no surprise that GWs, once detected, may be used to test some of the fundamental predictions of GR (and alternative theories of gravity). 

Evolution of a compact binary system has three phases: adiabatic inspiral, merger and ringdown. The {\it inspiral} phase is characterized by the adiabatic approximation ${{\dot\omega}\over\omega^2}\ll1$ where $\omega$ is the frequency of the binary and ${\dot\omega}$ the rate of change of frequency under GW radiation reaction. This phase can be analytically modelled using post-Newtonian (PN) approximation to GR and gravitational waveforms can be calculated with very good accuracy (see Ref.~\cite{Bliving} for a review of PN theory). The highly nonlinear merger phase, where the two compact objects merge to form a single compact object (most likely a BH), can only be modelled through high accuracy numerical relativity (see Ref.~\cite{Pretorius07Review,CentrellaNRrev10} for a review of numerical relativity in the context of GWs.). The newly formed compact object loses its asymmetries and settles down to an axially symmetric geometry emitting {\it ringdown} gravitational waveforms. This phase of the binary evolution can be modelled very well using BH perturbation theory (see Ref.~\cite{TSLivRev03,BertiQNMRev09}.).

The data analysis of coalescing compact binary signals are based on the technique of {\it matched filtering} (see Ref.~\cite{JKLivRev12} for a review of the GW data analysis in the case of Gaussian noise.). In matched filtering one cross-correlates the detector output with a bank of templates (which are copies of pre-calculated gravitational waveforms with different parameters) looking for correlations exceeding certain threshold indicating the presence of a signal.  As we just saw, all the three phases of binary evolution can be accurately modelled analytically or numerically using GR and waveforms can be computed. This prior computability of gravitational waveforms is central to the detection and parameter estimation of GWs from compact binaries.
%%%%%%%%%%%%%%%%%%%%%%%%%%%%%%%%%%%%%%%%%%%
\subsection{Classification of the tests}
%%%%%%%%%%%%%%%%%%%%%%%%%%%%%%%%%%%%%%%%%%%

There have been many proposals in the literature which discuss various types of tests that can be performed with GWs. These range from testing some particular nonlinear effects of GR\cite{BSat94,BSat95} to very generic tests of alternative theories of  gravity under a unified frame work~\cite{AIQS06a,AIQS06b,YunesPretorius09}. We wish to (broadly) classify various proposals to test GR and alternative theories of gravity into two categories. One where the GW data is compared against a particular model or theory of gravity, which we call {\it model dependent tests} and the second, which are {\it model independent} which means the GW data is compared against a generic waveform model with free parameters which are different for different theories of gravity which can be constrained from GW observations. For example, proposals to bound the Brans-Dicke parameter using GW observations (e.g. in Refs~\cite{WillBD94,KKS95}) would fall under the model dependent test category whereas parametrised tests of PN theory~\cite{AIQS06a,AIQS06b} and parametrised post-Einsteinian framework~\cite{YunesPretorius09} would belong to the model independent tests category. Before we go on and discuss these proposals, it is good to have an idea of the
PN waveforms, that are used and also about various detector configurations which are already constructed, undergoing upgrades or being proposed. These
two aspects are discussed in the next two subsections.
%%%%%%%%%%%%%%%%%%%%%%%%%%%%%%%%%%%%%%%%%%%%%%
\subsection{PN waveforms from inspiralling compact binaries in circular orbit}\label{sec:PN}
%%%%%%%%%%%%%%%%%%%%

Using PN approximation to GR, the response of the gravitational wave detector to an incident radiation from a source at a luminosity distance $D_L$ is given by
\begin{eqnarray}
h(t) &=& \frac{4{\cal C{\cal M}}}{D_L} \left [ \pi{\cal M}F(t) \right ]^{2/3} 
\cos \Phi(t),
\end{eqnarray}
where ${\cal M}=\eta^{3/5}M$ is the called  the {\it chirp mass} of
the system ($M$ being the total mass $m_1+m_2$ and $\eta={m_1m_2\over M^2}$ the symmetric mass ratio), $F(t)\equiv ({2\pi})^{-1} {d\Phi(t)/dt}$ 
is the instantaneous frequency
of the radiation, $0 \le {\cal C} \le 1$ is a dimensionless 
geometric factor that depends on position and orientation of the source and its average over all orientations is $\overline{\cal C}= 2/5.$ This waveform is referred to as pattern averaged waveform within the {\it restricted} waveform approximation, where the PN corrections to the phase of the waveform are accounted to the highest accuracy keeping the amplitude of the waveform to be leading Newtonian. This may be contrasted with the amplitude corrected {\it full} waveform where
PN corrections to the amplitude are also included~\cite{ABIQ04,BFIS08,SinVecc00a,SinVecc00b,ChrisAnand06,ABFO08}. The phasing formula is currently available till 3.5PN (${ (v/c)^7}$) beyond the leading quadrupolar order~\cite{BDIWW95,BFIJ02,BDEI04} for nonspinning binaries moving in circular orbits.

Since matched filtering is usually performed in the frequency domain, it is convenient to obtain the Fourier transform of the waveform above, which is efficiently done using stationary phase approximation. The frequency domain waveforms read
\begin{equation}
\tilde h(f) = {\cal A}\, f^{-7/6} \exp\left [ 2i\Psi(f) + i \frac{\pi}{4}\right ],
\end{equation}
with the Fourier amplitude ${\cal A}$ and phase $\Psi(f)$ given by
\begin{eqnarray} 
{\cal A} & = & \frac{\cal C}{D_L\pi^{2/3}} \sqrt{\frac{5}{24}} {\cal M}^{5/6},\nonumber\\
\Psi(f) & = & 2\pi f t_c + \Phi_c + \sum_k \left[\psi_k+\psi_{ kl} \ln f\right] f^{(k-5)/3}.
\label{eq:Fourier Phase}
\end{eqnarray} 
Here $t_c$ and $\Phi_c$ are the fiducial epoch of merger and the
phase of the signal at that epoch.  There are eight PN coefficients that appear in a 3.5PN phasing formula (including the logarithmic terms), which are given by~\cite{DIS01,DIS02,AISS05}:
\begin{eqnarray}
\label{eq:psikvsmass1}
\psi_k & = & \frac{3}{256\,\eta}(\pi\, M)^{(k-5)/3}\alpha_k,\\
\psi_{ kl} &=& \frac{3}{256\,\eta}(\pi\, M)^{(k-5)/3}\alpha_{kl},
\end{eqnarray}
where explicit expressions for $\alpha_k$  and $\alpha_{ kl} $ can be found, for example, in Eq.~(4)-(13) of Ref.~\cite{PaiArun12} (which is a slightly modified form of the same given in Refs.~\cite{AISS05,MAIS10}.).
 
Note that the above waveform is specialized to a
class of sources which move in circular orbits and have no spins. Though the first assumption is somewhat general (since gravitational wave radiation reaction is known to circularize the orbit of the binary) second assumption of neglect of spins is less general. Waveforms which account for the spins and spin induced precession can be found, for example, in Ref.~\cite{ABFO08}. Discussion of eccentric orbit binaries can be found in Ref.~\cite{GI97,DGI04,ABIQ07,ABIQ07tail,ABIS08}.

%%%%%%%%%%%%%%%%%%%%%%%%%%%%%%%%%%%%%%%%%%%%%%%%%
\subsection{Various GW detector configurations}
%%%%%%%%%%%%%%%%%%%%%%%%%%%%%%%%%%%%%%%%%%%%%%%
There are different interferometric GW detector configurations, some are already operational and undergoing upgrades and some are at the
level of proposals whose design studies are going on. In the paper we will consider advanced LIGO (aLIGO)\footnote{LIGO is an acronym for Laser Interferometric Gravitational wave Observatory.} as a representative
of the second generation ground-based interferometric GW detectors. (There are other detectors like advanced Virgo and KAGRA which are 
under upgrade or construction.) We will also consider a proposed 3rd generation ground-based interferometer called Einstein Telescope (ET)\cite{ET}.
For space-based detectors, we will consider the original LISA\footnote{LISA stands for Laser Interferometer Space Antenna.} configuration which has recently been downscaled to eLISA/NGO~\cite{eLISA}. These are still
at the level of design studies and a pathfinder mission for LISA is approaching its completion expecting to be launched in 2015~\cite{McNamara12}.
We will also consider the proposed space mission DECIGO for which design studies are being carried out (see the review by Kent Yagi in this issue~\cite{YagiRev12}). 

The ground-based detectors are sensitive to
high frequency GWs in the frequency range $1-10^4$ Hz, where as LISA/eLISA will be sensitive in the low frequency band $10^{-4}-0.1$ Hz.
DECIGO will have the sensitivity to GWs with frequency between $10^{-2}-100$Hz (hence the name Deci Hz Interferometric Gravitational wave Observatory).
Together these detectors span a frequency range of 8 orders of magnitude between $10^{-4}-10^4$ Hz.

In addition to these detectors there are space-based detectors like ASTROD-GW\footnote{ASTROD is an acronym for Astrodynamical Space Test of Relativity using Optical Devices}which are currently in the initial stages of its design studies~\cite{Ni2011}.
ASTROD-GW will be sensitive to the frequency band between $100$ nHz-1mHz. The potential of ASTROD-GW to test GR and alternative theories of gravity
needs to be explored in detail.

This article is organized in the following way. Sec.~\ref{sec:modeldept} reviews some of the model independent tests of theories of gravity which includes the scalar-tensor theories, massive graviton theories and Dynamical Chern-Simons theory. Proposed generic tests of gravity are reviewed in Sec.~\ref{sec:modelindept}. These include the Parametrised Post-Newtonian Tests, Parametrised Post Einsteinian framework and Bayesian methods of performing generic tests of strong field gravity. Impact of singular value decomposition for these tests are demonstrated in the context of eLISA in Sec.~\ref{sec:SVD}. Conclusions are given in Sec.~\ref{sec:conclusion}.

%%%%%%%%%%%%%%%%%%%%%%%%%%%%%%%%%%%%%%%%%%%%%%%%%%%%%%%%%%%%%%%%%%%%%%%%%%%%
\section{Model dependent tests of alternative theories}\label{sec:modeldept}
%%%%%%%%%%%%%%%%%%%%%%%%%%%%%%%%%%%%%%%%%%%%%%%%%%%%%%%%%%%%%%%%%%%%%%%%%%%
In this section, we discuss three important classes of alternative theories: scalar-tensor theories, massive graviton theories and dynamical Chern-Simons theory. We review literature and summarize the expected constraints on these theories from the future ground-based and space-based GW detectors.
\subsection{Scalar-Tensor theories}
Perhaps the simplest and most popular alternative theories of gravity are the scalar-tensor theories. In such theories, gravitational interaction is mediated not only by the metric tensor but also a scalar field which couples universally to matter. One well-studied example of scalar-tensor theories is 
Brans-Dicke (BD) theory\cite{BransDicke61}. It will be interesting to know what kind of bounds can be obtained on the scalar-tensor type theories from GW observations. In the subsections below, we discuss the various proposals in the literature discussing the possible bounds on scalar-tensor theories. The distinct feature of scalar-tensor theories is that they predict dipolar gravitational radiation
as opposed to GR where the leading order gravitational radiation is at quadrupolar order. This would modify the GW phasing formula where a dipolar term will appear. One can ask the question how well can GW observations constrain this dipolar term.
\subsubsection{Possible bounds on Brans-Dicke theory}
Will\cite{WillBD94} argued that the relative importance of the scalar field, in the gravitational phasing formula,  can be written down in terms of a   Brans-Dicke parameter $\omega_{\rm BD}$~\footnote{This parameter can be a function of the scalar field for generic scalar-tensor theories.}. General Relativity is recovered in the limit of large $\omega_{\rm BD}$ ($\omega_{\rm BD}\rightarrow \infty$). The important difference in gravitational wave emission within this theory is the presence of {\it dipolar} gravitational radiation. Will~\cite{WillBD94} argued that the PN phasing formula within the theory can be written parametrically introducing a (leading order) dipolar term in the phasing formula (see Eq. (3.19) of \cite{WillBD94}). 

\begin{figure}[h]
\begin{center}
\includegraphics[scale=0.3]{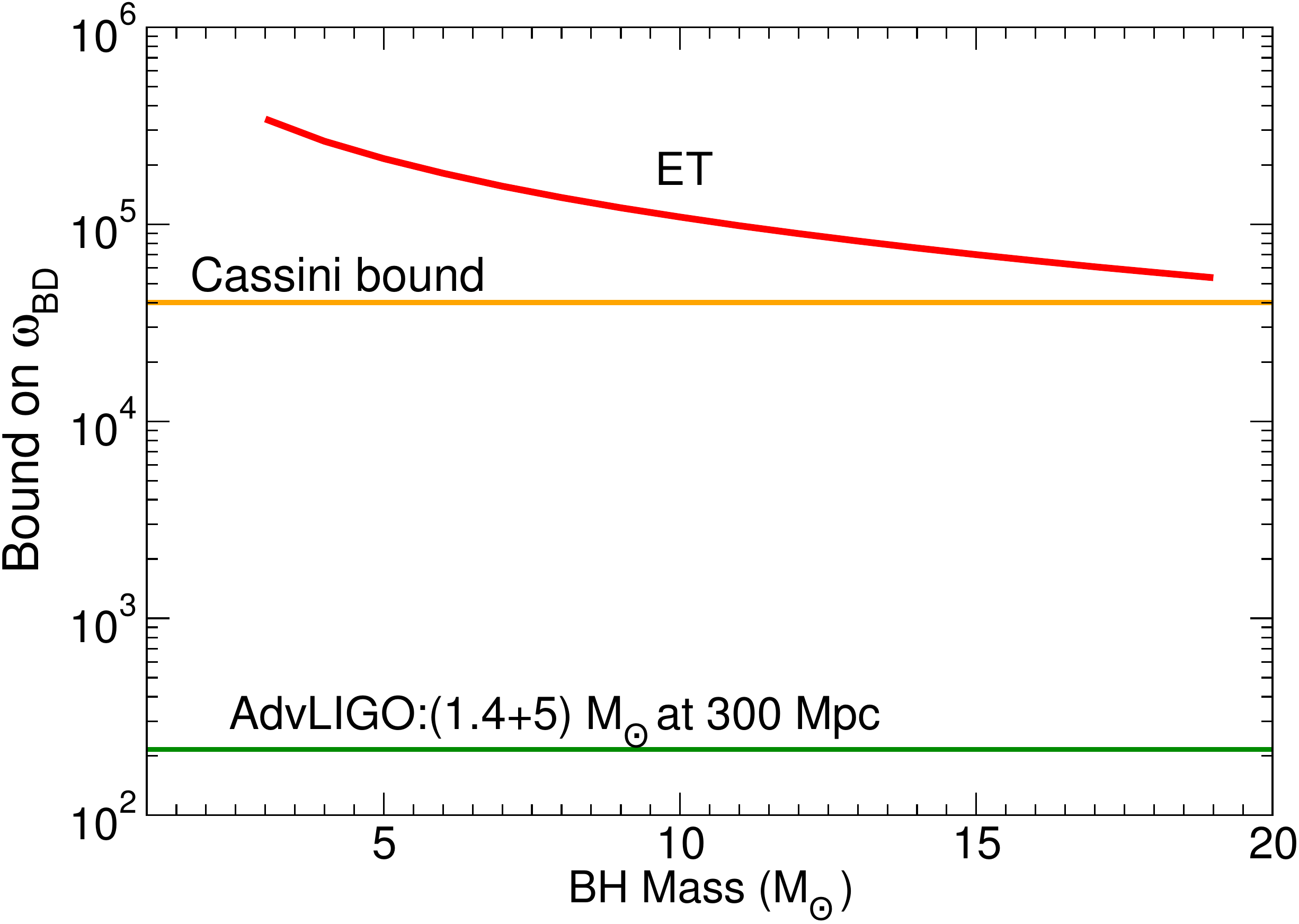}
\end{center}
{\caption{Projected bounds on $\omega_{\rm D}$ from the proposed Einstein Telescope for a NS-BH system as a function of BH mass (NS mass is assumed to be $1.4M_\odot$). These are compared against the best bounds that you expect from aLIGO and also the existing solar system bound from Cassini experiment. (Figure is taken from \cite{ArunET09}. )}\label{OmegaBD-ET}}
\end{figure}

The practical difficulty to obtain a good bound on the BD parameter is that gravitational waveforms of BD theory and GR are identical for binary BH systems. Further, for binary NS (BNS) systems, the dipolar term is proportional to the square of the difference in sensitivity parameters\footnote{Sensitivity parameter is a measure of self-gravitational binding energy per unit mass (see Sec. 3.3 of Ref.~\cite{Wthexp})} which is 0 for a binary BH (which is a consequence of the {\it no hair} theorem). For binary NSs the sensitivity parameter can be between $\sim 0.05-0.1$ and for NS-BH systems $\sim 0.3$. Thus NS-BH systems are the most interesting candidates to bound $\omega_{\rm BD}$. Using the matched filtering procedure, one can obtain the bounds on this parameter from GW observations of NS-BH systems. Fig.~\ref{OmegaBD-ET} compares the projected bounds on $\omega_{\rm BD}$ from aLIGO and proposed ET detectors. The best bounds  $\sim 10^5$ would come from a NS-BH systems where mass of the BH is less than $5M_{\odot}$. We also note that if ET can observe binary NSs with masses $(1.4+2)M_{\odot}$ with a sensitivity parameter $=0.1$, the projected bound on ${\omega_{\rm BD}}\sim 10^4$, if the source is at  100 Mpc.

LISA may be able to yield interesting constraints on the BD parameter as it is sensitive to NS-BH binaries with BH mass $\sim 400-10^3M_{\odot}$, if such systems are close-by (at a distance of $\sim$ 20 Mpc, which will have an SNR of $\sim 10$) in which case it may yield a bound comparable
to or better than Cassini measurements~\cite{BBW05a}.  Our calculations, similar to Berti et al.\cite{BBW05a}, show that bounds expected on $\omega_{\rm BD}$ from eLISA
could be worse by a factor ranging between 3 (for $1-400M_{\odot}$) to 6 (for $(1-10^3M_{\odot}$). On an average the bounds expected from eLISA would be $\sim10^{4}$.%\footnote{We thank Wei-Tou Ni for suggesting to include the eLISA numbers.}

The best bet for bounding BD parameter could be the 
future probe like DECIGO which has sensitivity in the frequency range $10^{-2}-10^{2}$ Hz which can observe NS-BH binaries where the BH is of stellar mass range ($\sim 10M_{\odot}$). Sensitivity to the low frequency GWs as well as the stellar mass range
of sources make DECIGO the ideal GW detector to yield constraints as high as $\sim 10^6$ as found in Ref.~\cite{YagiTanaka09b}. They further argue that if they fold in the effect of population of sources DECIGO can observe (roughly ${\cal O}(10^4)$), the bounds can become as high as $10^8$.

Table.~\ref{TableBD} summarizes the expected bounds from aLIGO, ET, eLISA, LISA and DECIGO.
\begin{center}
\begin{table}
\tbl{Projected bounds on $\omega_{\rm BD}$ 
for various detector configurations. 
These are compared against the already existing solar system bounds from the Cassini experiment\cite{Bertotti03}. Note that an SNR of 10 for $1-400M_{\odot}$  for LISA corresponds to distance of $\sim 20$ Mpc (and even less for eLISA). For ET, a seismic low frequency cut-off of 1Hz is assumed.
}
{\begin{tabular}{ccccc}
\hline\hline
Detector & System &  Specification & Expected bound on $\omega_{\rm BD}$ & Reference\\
\hline\hline
aLIGO & (1.4+5)$M_{\odot}$ & 100 Mpc & $\sim$ 100 & Arun in \cite{ArunET09}\\
ET & (1.4+5)$M_{\odot}$ & 100 Mpc & $\sim10^5$  & Arun in\cite{ArunET09}\\
ET & (1.4+2)$M_{\odot}$ (BNS) & 100 Mpc & $\sim10^4$  & This paper.\\
eLISA & $(1.4+400) M_{\odot}$ & SNR=10 & $\sim10^4$ & This paper.\\
LISA & $(1.4+400) M_{\odot}$ & SNR=10 & $\sim10^5$ & Berti et al\cite{BBW05a}\\
DECIGO & $(1.4+10)M_{\odot}$ & SNR=10 & $\sim10^6$ & Yagi \& Tanaka\cite{YagiTanaka09b}\\
\hline
Cassini & Solar system experiment& & $10^4$ & Bertotti et al\cite{Bertotti03}\\
\hline\hline
\end{tabular}\label{TableBD}}
\end{table}
\end{center}

Currently, the best bound that exists for the parameter $\omega_{\rm BD}\sim 10^4$ is from {\tt Cassini} which is a solar system experiment\cite{Bertotti03}. From the Table it is clear that advanced third generation detectors and space based detectors like LISA would be the best bets to beat the solar system constraints on the Brans-Dicke parameter.

Many groups have worked on understanding various features of the waveform that may affect the bound on $\omega_{\rm BD}$. This includes the effect of LISA noise PSD\cite{WillYunes04}, effect of nonprecessing spins~\cite{BBW05a, YagiTanaka09b} and effect of orbital eccentricity~\cite{YagiTanaka09a}.

Some of the upcoming/proposed solar system experiments are capable of measuring the time delay parameter $\gamma$ to unprecedented accuracies~\cite{
ASTROD2012}. For
example ASTROD-I mission may be able to bound $\gamma$ at the level of $3\times10^{-8}$ which is three orders of magnitude better than the present bound.
This will also lead to very accurate constraints on BD parameter, albeit, from non-GW observations. 
%%%%%%%%%%%%%%%%%%%%%%%%%%%%%%%%%%%%%%%%%%%%%%%%%%%%%%%%%%%%%%%%%%%
\subsubsection{Possible bounds on generic scalar-tensor theories}
%%%%%%%%%%%%%%%%%%%%%%%%%%%%%%%%%%%%%%%%%%%%%%%%%%%%%%%%%%%%%%%%%%%%
There have been studies about the bounds that GW observations can put on generic scalar-tensor theories. Damour and Esposito-Farese compared the probing power of (ground-based) GW observations and binary pulsar observations in the context of a two-parameter family of scalar-tensor theories\cite{Dgef98} and concluded that binary pulsar observations are better probes of strong-field radiative aspects of relativistic gravity.

 More recently, the possible bounds on generic dipolar radiation was studied in Ref.\cite{Arun2012}, where
the bounds on a dipolar phasing parameter $\beta$ (which is related to $\omega_{\rm BD}$ in the case of BD theory) and an amplitude parameter $\alpha$ was obtained for aLIGO and ET configurations. A parametrised gravitational waveform in Fourier domain (given in Eq.~(6) and (7) of that paper) forms the basis of their analysis. Here based on the same formalism we extend the results to the case of space-based detectors, especially eLISA and LISA detectors.

\begin{figure}[h]
\begin{center}
\includegraphics[scale=0.3]{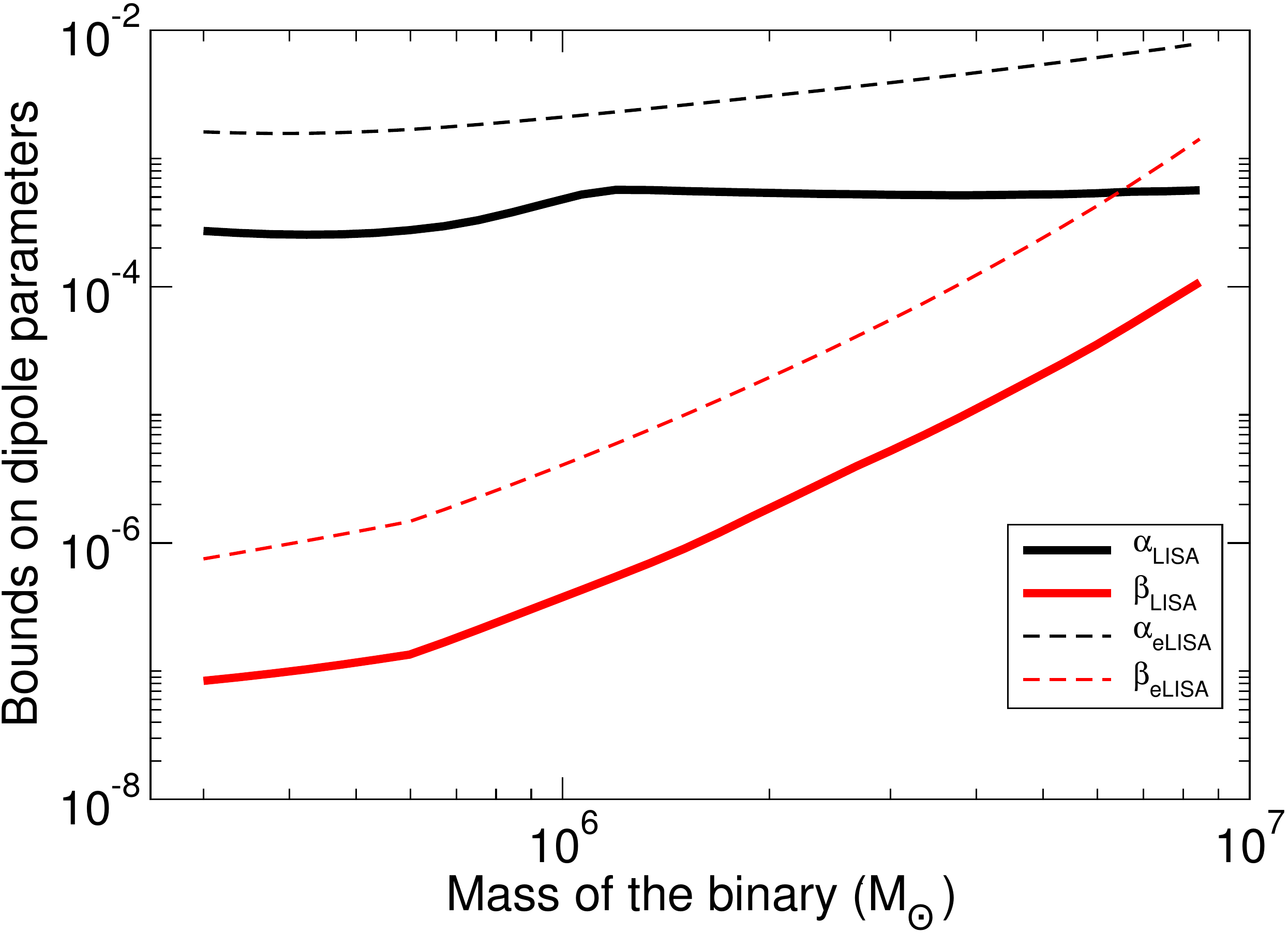}
%\label{LISA-eLISA}
\end{center}
{\caption{Projected bounds on $\alpha$ and $\beta$, describing generic dipolar gravitational radiation using (classic)LISA and eLISA detectors, as a function of the mass of the binary. Sources are assumed to be at a distance of 3 Gpc, irrespective of the mass. The errors were obtained using Fisher matrix calculation.}\label{LISA-eLISA}}
\end{figure}

 Fig.~\ref{LISA-eLISA} presents
the bounds on $\alpha$ and $\beta$ expected from eLISA and LISA configurations.
The typical bounds from various detectors are compared in Table~\ref{Alpha-Beta}. The table indicates an order of magnitude improvement in the bound on the parameters using classic LISA configuration, whereas the bounds from ET and eLISA are comparable. It should be pointed out that the estimates are obtained using Fisher matrix formalism which is valid in the limit of high SNR.
\begin{center}
\begin{table}
\tbl{Projected bounds on $\alpha$ and $\beta$ parameters describing generic dipole GW radiation\cite{Arun2012}. For ground-based detectors, sources are assumed to be at 300 Mpc and for eLISA and LISA sources are assumed to be at 3 Gpc.}
{\begin{tabular}{ccc}
\hline\hline
 & $\alpha$ & $\beta$\\
\hline\hline\\
aLIGO & $10^{-2}$ & $10^{-5}$\\
ET & $10^{-3}$ & $10^{-6}$\\
eLISA & $10^{-3}$ & $10^{-6}$\\
LISA & $10^{-4}$ & $10^{-7}$\\
\hline\hline
\end{tabular}\label{Alpha-Beta}
}
\end{table}
\end{center}

More recently, Berti et al.\cite{BertiLightScalar12} studied possible bounds on the $\omega_{\rm BD}$-like coupling parameter if the scalar field has a small mass $m_s$. They showed that compact binary observations can bound the combination ${m_s\over\sqrt{\omega_{\rm BD}}}$. They obtained $(\frac{m_s}{\sqrt{\omega_{\rm BD}}})(\frac{\rho}{10})\leq 10^{-15}, 10^{-16}, 10^{-19}$ eV for aLIGO, ET and eLISA, respectively, where $\rho$ denotes the signal to noise ratio of the observation.

We would like to point out that in our eLISA calculations, we have assumed the
upper cut-off frequency of eLISA to be 0.1Hz. The bounds on various scalar tensor parameters obtained this way would be worse than those obtained assuming an upper cut-off frequency of 1Hz. While comparing the eLISA numbers of this paper with that of Yagi~\cite{YagiRev12}, it should be kept in mind that Yagi has assumed the upper cut-off frequency for eLISA to be 1Hz and hence the bounds he obtained would be better than those reported in this paper.
\subsection{Possible bounds on Dynamical Chern-Simons Gravity}
Chern-Simons (CS) modified gravity theory is one of the most interesting extensions of GR where gravitational field is coupled with a scalar field through a parity-violating CS parameter (see \cite{CSreview09} for a review). In the dynamical version of the theory, the scalar field is treated as a dynamical field. This type of modification naturally appears in string theory and loop quantum gravity. Constraints on dynamical CS parameter is rather weak because it interacts only with gravity and it reduces to GR in the weak-field limit and has the same post-Newtonian parameters as GR~\cite{AlexanderYunes07a,AlexanderYunes07b}. Recently Yagi et al. obtained the leading correction to the GR phasing formula from CS type
modification\cite{YYT12}. This term occurs at 2PN and they estimated the expected bounds on this parameter from GW observations. The CS parameter $\zeta^{1/4}$scales as distance and it tends to GR in the limit $\zeta\rightarrow0$. Hence
the smaller the bound on $\zeta$, the better. They estimated typical bounds of ${\cal O}(10-100)$kms with advanced ground-based detectors which are better than the expected bound ${\cal O}(10^5-10^6)$kms from LISA. This is because the expected bounds are of the order of the length scale of the target system which in turn is ${\cal O}(M)$ where $M$ is the total mass of the system. Since LISA is sensitive to super massive black holes, the bounds are weaker. Its worth noting that the expected bounds from ground-based and space-based GW detectors are 6-7 orders of magnitude stronger than that from solar system experiment, which is $\xi^{1/4}<{\cal O}(10^8)$kms~\cite{DCSSolar11}.
\subsection{Massive graviton theories}
Massive graviton (MG) theories are a class of theories where gravitation
is propagated by a massive field as a result of which GWs travel with velocities different from the speed of light unlike GR. In this case the velocity of propagation will depend up on their frequency as $(\frac{v_g}{c})^2=1-(\frac{c}{\lambda_g})^2$
and the effective Newtonian potential may be modelled to be of the Yukawa form $\propto r^{-1} e^{-\frac{r}{\lambda_g}}$ where $\lambda_g$ is the gravitational compton wavelength. Will\cite{Will98} argued that, if graviton is massive, it will result in different propagation speeds of the low frequency and high frequency GW components of an inspiral waveform resulting a modification of the GW phasing predicted by GR. This modification can be parametrised in terms of $\lambda_g$ which hence can be bounded from GW observations which translates in to bounding the mass of the graviton.
General Relativity is recovered, obviously, in the limit $\lambda_g\rightarrow \infty$.

Many subsequent works improved this model incorporating various features of the waveform. Effect of (non-precessing) spins was studied in the context of space based LISA detector by Berti et al\cite{BBW05a}. Effect of spin precession was investigated in detail by Stavridis and Will~\cite{SW09} and also by Yagi and Tanaka~\cite{YagiTanaka09a}.
It was found that the nonprecessing spins deteriorate the bounds due to large correlation of the massive graviton parameter with the spin parameters.
But spin induced precession and the resulting modulations of the waveform makes up for the deterioration. Effect of amplitude corrected PN waveforms
was studied by Arun and Will\cite{AW09} and they found an improvement roughly by an order of magnitude, for ground-based and space-based detectors
for high mass sources (see Fig. 1 of Ref.\cite{AW09}). Our calculation, following Ref.~\cite{AW09}, shows that eLISA is expected to yield bounds on $\lambda_g$ which is $\sim10^{15}$ kms. The use of FWF can improve the bound by a factor of 7 for high mass systems for a mass ratio of 10. Recently Huwyler et al. studied the effect of incorporating higher harmonics {\it and} spin precession in constraining alternative theories of gravity. They found, for precessing case, higher harmonics results in an improvement $1.6$ times compared to the restricted waveforms~\cite{HKJ11}.
\begin{figure}[h]
\begin{center}
\includegraphics[scale=0.3]{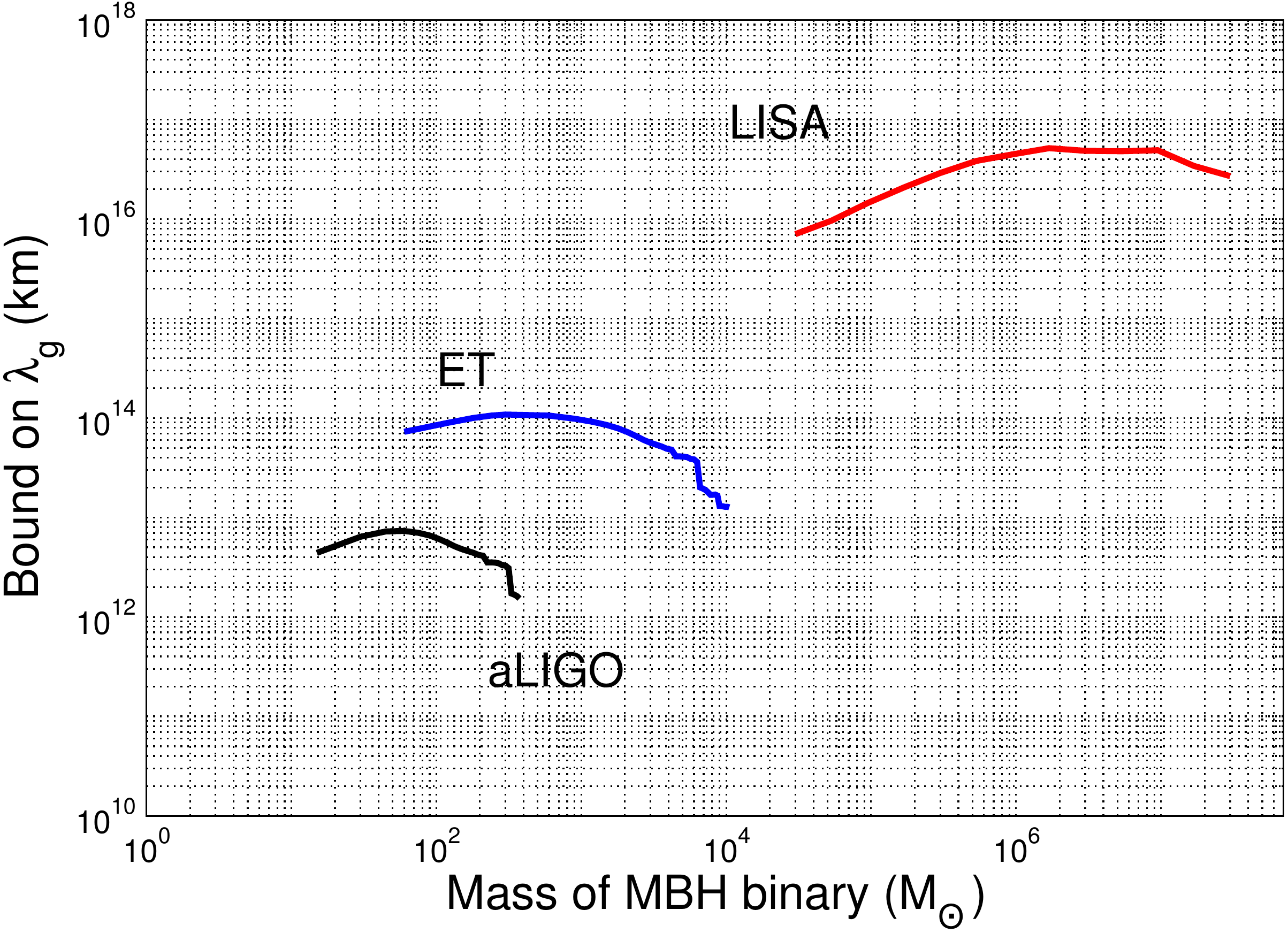}
\end{center}
{\caption{Comparison of expected bounds for aLIGO, ET and LISA detectors using amplitude corrected full waveform. Figure taken from Ref.\cite{AW09}. Bounds roughly improve by an order of magnitude in going from aLIGO to ET and ET to LISA.}\label{PlotMG}}
\end{figure}

Effect of orbital eccentricity on the estimation of the massive graviton parameter was investigated by Yagi and Tanaka~\cite{YagiTanaka09a}. They found that including (nonprecessing) spins and orbital eccentricity results in deterioration of the
bound where as invoking (simple) precession improves the bound almost by an order of magnitude. Keppel and Ajith\cite{KeppelAjith10} studied the effect of incorporating the merger and ringdown phases of the binary BH waveforms using a phenomenological parametrisation of the numerical relativity waveforms obtained in Ref.~\cite{AjithNR07b}. They argued that the inclusion of the merger and ringdown parts can lead to drastic improvement for the ground-based and space-based detectors. Berti et al. studied~\cite{BGS11} the effect of population on the bounds on $\lambda_g$ and found that that the bounds improve proportional to the square root of number
of observations.
The  best expected bounds for different detector configurations and different waveform models are summarized in Table~\ref{MG} below. The best bound on $\lambda_g$ would be $\sim 10^{13}, 10^{14}\; {\rm and}\; 10^{17}$ kms for aLIGO, ET and LISA, respectively. Recently Mirshekari et al.\cite{MYW11} proposed a generalization of this idea to capture any generic Lorentz invariance violation which leads to modified dispersion relation and discussed the possible constraints from future GW observations. Formulation of the problem from a Bayesian view point was presented in Ref.~\cite{BayesianMG11} by Del Pozzo et al.
This method is very handy while implementing the scheme directly in the parameter estimation pipelines.

\begin{table}
\tbl{Expected `best' bounds on $\lambda_g$ (in kms) for various detector configurations and waveform features. FWF refers to the amplitude corrected PN waveform\cite{BIWW96,ABIQ04,BFIS08,ChrisAnand06b} (without spins) and IMR is the phenomenological inspiral-merger-ringdown waveform obtained by Ajith et al~\cite{AjithNR07b}. For ground-based detectors aLIGO and ET sources are assumed to be at 300 Mpc and for eLISA, LISA and DECIGO, sources are assumed to be at 3 Gpc. Effects of spin, eccentricity etc cannot be read-off from the table as we have taken the best possible bounds for each waveform model.}
{\begin{tabular}{ccccccc}
\hline\hline
 & non-spinning & FWF & nonprecessing & precessing &eccentricity& IMR\\
\hline\\
aLIGO&$10^{12}$(Ref.\cite{Will98}) & $10^{13}$(Ref.~\cite{AW09}) & -&- &-&$10^{13}$(Ref.~\cite{KeppelAjith10}) \\
ET & $10^{13}$(Ref.\cite{AW09})& $10^{14}$(Ref.~\cite{AW09}) & -&- &-& $10^{14}$(Ref.~\cite{KeppelAjith10})\\
eLISA &$10^{15}$(This paper.) & $10^{15}$(This paper.) & - &-&-&- \\
LISA &$10^{16}$(Ref.\cite{BBW05a}) & $10^{16}$(Ref.~\cite{AW09}) & $10^{16}$(Ref.~\cite{BBW05a})&$10^{16}$(Ref.~\cite{SW09,YagiTanaka09a})&$10^{16}$~(Ref.~\cite{YagiTanaka09a})& $10^{17}$(Ref.~\cite{KeppelAjith10}) \\
DECIGO & $10^{15}$ (Ref.\cite{YagiTanaka09b})&- & $10^{15}$ (Ref.~\cite{YagiTanaka09b})&-& $10^{15}$ (Ref.~\cite{YagiTanaka09b})&-\\
\hline\hline
\end{tabular}\label{MG}
}
\end{table}

%%%%%%%%%%%%%%%%%%%%%%%%%%%%%%%%%%%%%%%%%%%%%%%%%%%%%%%%%%%%%%%%%%%%%
\section{Model independent Tests}\label{sec:modelindept}
Having discussed various model-dependent constraints that may be obtained on various alternative theories of gravity, this section is devoted to discussing generic, model independent tests of GR and alternative theories of gravity. 
We discuss the proposals in chronological order.
%%%%%%%%%%%%%%%%%%%%%%%%%%%%%%%%%%%%%%%%%%%%%%%%%%%%%%%%%%%%%%
\subsection{Parametrised Tests of Post-Newtonian Theory (PTPN)}\label{sec:PTPN}
%%%%%%%%%%%%%%%%%%%%%%%%%%%%%%%%%%%%%%%%%%%%%%%%%%%%%%%%%%%%%%%
As described in Sec.~\ref{sec:PN}, 3.5PN phasing formula for nonspinning binaries in circular orbit is characterized by eight PN coefficients which are predictions of PN approximation to GR. An alternative theory would, in principle, predict a different set of phasing coefficients. Hence, accurate measurements of these phasing coefficients can be interpreted as tests of PN approximation to GR. Thus if one uses a parametrised phasing formula, where every term in the PN phasing formula is treated as an independent parameter, GW observations can determine these free parameters (within the observational error bars) leading to a test of GR\cite{AIQS06a}. Since eight of these coefficients are fully characterised by two component masses, one can perform a consistency test between these PN coefficients in the mass plane of the binary (see Fig. 2 of Ref~\cite{AIQS06a}).

However, it turns out that there are large degeneracies between these parameters which are essentially functions of the two component masses of the binary. We are actually using eight parameters to fit a data which is actually characterised by two parameters leading to strong correlations among the parameters~\cite{AIQS06a}. However, within these caveats, Ref.\cite{AIQS06a} showed that for a narrow range of masses between $10^5-10^6M_{\odot}$ in the LISA band, all of the eight PN parameters can be estimated with ($1-\sigma$) relative errors $\leq 1$ and the first three PN parameters can be estimated with accuracies $\sim 10^{-3}, 10^{-2}, 10^{-1}$ respectively. As is obvious, though a clean procedure, it is not applicable to many sources especially those which ground-based detectors can observe.

This scheme was modified in Ref.~\cite{AIQS06b} where the authors proposed to
use only three (out of eight) parameters as independent, parametrising the rest
of the parameters in terms of any of the two PN parameters and perform consistency tests in the mass plane of the binary\footnote{Since two phasing coefficients are necessary to determine the mass, we need at least three phasing parameters to perform a consistency test.}. Advantage of this scheme is that we are dealing with only three free parameters which drastically improves the effectiveness of the test. The best choice would be to use the first two PN coefficients as basic parameters to be used for parametrising all other PN terms except one which may be treated as a test parameter. Keeping the basic parameters the same, one can change the test parameters yielding different tests (not all are independent). They estimated that all the 6 higher order PN parameters, when used a test parameters, can be estimated with $1-\sigma$ relative errors less than unity for a mass range $1-10M_{\odot}$ for ET-type third generation detectors. Almost for the entire mass range, LISA can determine them with much better (1-2 orders of magnitude) accuracies, the relative accuracies being $\sim 10^{-2}$ for $10^5-10^6M_{\odot}$(see Fig.~2 of Ref.\cite{AIQS06b}). However, with aLIGO not all of these test parameters can be estimated with good accuracy.
Mishra et al. studied the effect of using amplitude corrected PN waveforms on this proposal~\cite{MAIS10}. Employing the best available inspiral waveforms with 3.5PN phase and 3PN amplitude, they showed that errors on the phasing parameters, especially at high masses, are reduced because of the use of amplitude corrected waveforms. While aLIGO can test only up to 1.5PN term in the phasing, ET can test the consistency between various PN coefficients over the mass range $11-44M_{\odot}$. Effect of the seismic cut-off frequency on these errors and the systematic bias due to spins were also studied.

Recently Pai and Arun~\cite{PaiArun12} showed that starting from the proposal in Ref.~\cite{AIQS06a}, use of singular value decomposition can help one obtain the most dominant three new parameters which accounts for most of the information. This method of finding the three new parameters is more natural. This also brings in aLIGO into the picture for which all of these three {\it new} parameters can be estimated with a $1-\sigma$ relative errors $10^{-4}-10^{-2}$ for low mass systems.
Accuracies are even more spectacular ($\sim10^{-5}$) for low mass binaries in the ET band. Extension of this proposal to estimate the bounds on the new phasing parameters for eLISA and LISA is discussed in Sec.~\ref{sec:SVD} below.
%%%%%%%%%%%%%%%%%%%%%%%%%%%%%%%%%%%%%%%%%%%%%%%%%%%%%
\subsection{Parametrised post-Einsteinian Formalism (PPE)}
%%%%%%%%%%%%%%%%%%%%%%%%%%%%%%%%%%%%%%%%%%%%%%%%%%%%%
Parametrised post-Einsteinian framework~\cite{YunesPretorius09} is developed with the goal of quantifying the theoretical bias due to the use of general relativistic waveforms in data analysis for GWs. The authors wrote down a generic parametrised waveform which characterizes the departures from GR through the free parameters which are present in the waveform and have different values in different theories. It can be considered to be a generalization of PTPN framework to consider more general waveforms which may have a different PN structure from GR (in terms of the exponents of the frequency). Needless to say, this reduces to PTPN for theories which have a PN structure similar to that of GR. They also account for the contribution from merger and ringdown phases of the binary evolution. The formalism is, in principle, applicable to metric theories of gravity which have same weak-field limit as GR but deviate when the gravitational fields are strong.

The first detailed study to quantify the parameter biases using PPE was reported in Ref.~\cite{PPE2011} where they used Bayesian inference and model selection with the simulated data of advanced LIGO/Virgo network and also for the proposed space-based detector LISA. The qualitative conclusions they reached are the following. (1) GW observations can constrain various higher order PN coefficients much more tightly than binary pulsar observations. (2) Parameter estimates can have significant bias if GR templates are used to recover signals which are described by an alternative theory of gravity. (3) Detection efficiency of GR templates are also seriously affected if the data is characterized by a non-GR theory.

Further, Chatziioannou et al.~\cite{PPEPol2012} put forward a model-independent test of GR which permits constraints on the GW polarization contents from GW observations of inspiralling compact binaries. Considering three specific alternative theories of gravity, viz, BD theory, Rosen's bi-metric theory and Lightman-Lee theory, they computed the response function in the case of ground-based GW detectors (for the inspiral phase). These helped them to extend the PPE waveform to account for 6 GW polarization states, the maximum a metric theory of gravity can have. They also discussed various data analysis techniques which can lead to the bounds on the polarization parameters.
PPE and its extensions give a very general frame-work for writing down generic waveforms
and discussing possible constraints from GW observations. 

\subsection{Bayesian Inference and Tests of Gravity}
Bayesian approach towards parameter estimation of GW signals is a relatively new
field (see for example Ref.~\cite{VeitchVecchio09}). If we want to compare two theories of gravity given the observed data and all the priors that we have in hand, the first step is to compute the posterior probability of the two theories
using Bayes theorem. This leads to the computation of the important quantity {\it Odds ratio} which is nothing but the ratio of posterior probabilities of the two theories. Odds ratio is also the product of ratios of the priors of the two theories ({\it prior odds}) and the {\it Bayes factor}. 
The signal model selection, between GR and an alternative theory, 
%The detection of a signal from alternative theory of gravity 
is indicated by the corresponding Odds ratio exceeding a threshold value which in turn is based on the requirement of a given false alarm rate\footnote{False alarm rate, in the case where GR is compared against an alternative theory of gravity, would be the fraction of events for which Odds ratio is above the threshold, but the underlying theory is GR.}.
 The estimation of the unknown parameters of the theory is carried out by first computation of joint probability distribution function of the
parameters and then marginalising it over all the parameters except the parameter
of interest. Computation of these quantities is done via numerical simulations involving Monte Carlo techniques such as nested sampling~\cite{VeitchVecchio09}.

 In Ref.~\cite{BayesianMG11}, Del Pozzo et al. recast the problem of bounding the mass of the graviton from the point of view of Bayesian model selection and argued that his method can be extended to other alternative theories of gravity. A generic test of strong field gravity, within the Bayesian framework,
was proposed by Li et al.~\cite{LiEtal2011,LiEtalAmaldi2011}. In these works they used the proposal of Ref.~\cite{AIQS06b} (described earlier) to test the PN coefficients of GR recasting it using Bayesian model selection. They did studies of a few cases of deviation from GR, including the cases where phasing has a 1.25PN term (which means a a power of frequency which is different from GR). Using simulated data, they found that the framework was able to detect such deviations even if such deviations
are not accommodated in the model waveforms, provided the deviations 
near the most sensitive frequencies of the LIGO band are at least a few percent ($\sim 5$ rad at 150Hz.)

More recently Vallisneri put forward a simple formula, based on Bayesian inference, to characterize tests of GR~\cite{Vallisneri12}. He argued that, in the limit of large SNR and Gaussian noise, the detectability of an alternative theory can be fully characterized by a single quantity, viz, the Fitting Factor~\cite{Apos95} between the GR waveform and the family of alternative theory of gravity of our interest.
Based on his scheme, he found that aLIGO type second generation ground-based detectors may be able to measure deviations from GR waveforms which are $1-10\%$ (corresponding to Fitting factors between $0.9-0.99$).
%%%%%%%%%%%%%%%%%%%%%%%%%%%%%%%%%%%%%%%%%%%%%%%%%
\section{Singular Value Decomposition and parametrised tests of post-Newtonian theory in the context of eLISA}\label{sec:SVD}
%%%%%%%%%%%%%%%%%%%%%%%%%%%%%%%%%%%%%
 Recently Pai and Arun proposed~\cite{PaiArun12} the use of truncated singular value decomposition (SVD) technique to minimise the near-singular nature of the Fisher
matrix (arising due to strong correlations as seen in Ref.~\cite{AIQS06a} ) 
in the space of PN phasing coefficients. The truncated SVD approach in the context of parametrised tests of PN theory, helps in identifying the most dominant phasing parameters (which are linear combinations of the original PN phasing coefficients) which can be determined with good accuracy. In Ref.~\cite{PaiArun12} this method was described in the context of advanced ground-based detectors advanced LIGO and Einstein telescope detectors. In this section we extend this method to the case of space-based detectors like eLISA and LISA. 
%%%%%%%%%%%%%%%%%%%%%%%%%%%%%%%%%%%%%%%%%%%%%%%%%%%%%%%%%%%%%%%%%%%%%%%
\subsection{Singular Value Decomposition of Fisher Information Matrix}
%%%%%%%%%%%%%%%%%%%%%%%%%%%%%%%%%%%%%%%%%%%%%%%%%%%%%%%%%%%%%%%%%%%%%%
 The basic idea is discussed detail in Sec.~3 of \cite{PaiArun12}. We present a concise qualitative summary of the procedure here not entering too much into the mathematical aspects. Fisher information matrix is constructed by the noise weighted inner product of tangent vectors of gravitational waveform w.r.t various parameters that 
are treated as independent in the problem. In the presence of Gaussian noise and large SNR,
square root of the diagonal entries of the inverse of the Fisher matrix (covariance matrix) gives $1-\sigma$ errors on various parameters~\cite{CF94}. If the parameters describing the waveform are strongly correlated, it can lead it ill-conditioned Fisher matrix (see Appendix B of Ref.~\cite{BBW05a} for a discussion).
 
Truncated Singular Value Decomposition (SVD) is a technique that is widely used in data analysis for reducing the dimensionality of the data by transforming  correlated variables into a set of uncorrelated ones and understanding which of the newly constructed parameters are the most significant to describe the data and re-express the data in terms of these most significant ones. In our case, as we saw in Sec.~\ref{sec:PTPN},
there is strong correlation among the eight PN phasing parameters when all of the are treated as independent. We can apply the scheme of Truncated SVD to obtain the most significant parameters which are linear combination of the original PN coefficients. Below is a more technical details of this procedure.

%We start with an SVD of the noise-weighted tangent vector of the frequency domain gravitational waveform, $\frac{\partial{{\tilde h}(f)}}{\partial \theta^i}$ (where $\theta^i$ is one of the parameters describing the waveform). We showed that SVD of  $\frac{\partial{{\tilde h}(f)}}{\partial \theta^i}$ for all $i$ (eight in total) puts Fisher matrix in a diagonal form

We write the Fisher information matrix in the 8 dimensional PN phase parameter space, namely ${\bf \theta} \equiv \{\psi_0,\psi_2,\psi_3,\psi_5,\psi_{5l},\psi_6,\psi_{6l},\psi_7 \}$, in terms of $\theta^\alpha$ with $\alpha = 1 \ldots 8$ in the discrete frequency domain as
\begin{equation}
\Gamma_{ij} = 4 \Re \left[ \sum_{k=0}^{N-1} \frac{{\tilde{h}}_{,i}(f_k)}{\sqrt{S(f_k)}}
\frac{{\tilde{h}}^*_{,j}(f_k)}{\sqrt{S(f_k)}} \right]~\Delta f \,.
\end{equation}
where at $k$-th frequency bin $f_k$, ${\tilde{h}}_{,i}(f_k) = \partial{\tilde{h}}(f_k)/\partial \theta_i$ denotes the $i-th$ component of the tangent
vector of the GW signal in the intrinsic PN phase space and $\Delta f$ is the width of the frequency bin. We stack all the tangent vectors in a matrix form with column index representing the frequency index as 
${\bf H}=\{h_{ik} \equiv \frac{{\tilde{h}}_{,i}(f_k) \sqrt{\Delta f}}{\sqrt{S(f_k)}} \}_{8 \times N}$. 

Then the Fisher matrix takes a compact form in terms of ${\bf H}$ as
\begin{equation}
\Gamma = 2 {\bf H}{\bf H}^{\dagger},
\end{equation}
where the dagger ($\dagger$) denotes transposed conjugate of the matrix.

Though not very surprising result, this compact form of Fisher enables us to
treat the ill-conditioning of the Fisher which arises due to the correlations
between the parameters which was one of the main hurdles in the previous work~\cite{AIQS06a}. We note that ${\bf H}$ acts like a building block for the Fisher. It carries the information about the spectral 
content of the GW signal variation in the interferometer's noise in ${\bf \theta}$ space. The natural way to remove the correlations would be to look
for those directions in the parameter space where the signal variation is maximum. Technically, these are the principle directions or
principal components of the Fisher matrix. The SVD provides the natural way
to arrive at the principal components.  

The SVD of ${\bf H}$ is ${\bf H}={\bf U_H}{\bf \Omega_H}{\bf V_H}^{\dagger}$ where ${\bf U_H}_{ m \times m}$ and ${\bf V_H}_{n \times n}$ are left and 
right unitary square matrices 
and ${\bf \Omega_H}$ arranges the singular values in descending order as diagonal entries. The singular values govern the contribution of the
singular vector in a particular direction. Substituting the SVD of ${\bf H}$ in $\Gamma$, we obtain the SVD of $\Gamma$ as
\begin{equation}
\label{Gamma2}
{\bf \Gamma = U_H \Sigma\, U_H^{\dag}},
\end{equation}
where ${\bf \Sigma=2\Omega_H\Omega_H^{\dagger}}$.
%%%%%%%%%%%%%%%%%%%%%%%%%%%%%%%%%%%%%%%%%%%%%%%%%%%%%%%%%%%%%%%%%%%%%%%
\subsection{Truncated SVD, new phase parameters and relative errors}
%%%%%%%%%%%%%%%%%%%%%%%%%%%%%%%%%%%%%%%%%%%%%%%%%%%%%%%%%%%%%%%%%%%%%
For a near-singular matrix, some of the singular values are very small compared to the most dominant one i.e. very little information lies along those directions. In the SVD of ${\bf H}$, the signal variation along those directions are very small or negligible
% i.e. incorporating them is nothing but 
%adding zeros in construction of ${\bf H}$. 
which makes the matrix ill-conditioned. This ill-conditionedness in turn gives rise to unreliable estimation
of errors. We propose to truncate the singular matrix of ${\bf H}$ to the dominant singular values. If $r$ is the number of dominant
values, the Fisher matrix in terms of truncated SVD is
\begin{equation}
\label{Gamma3}
{\bf \Gamma = U^t_H \Sigma^t U_H^{t\dag}}\hspace{0.5in} {\Rightarrow}\hspace{0.5in} {\bf  U_H^{t\dag} \Gamma U_H^t} = {\bf \Sigma^t}
\end{equation}
where the index $t$ corresponds to the truncation. The ${\bf U_H^t}_{8 \times r}$ gives the singular-vectors corresponding to the dominant singular values. The the singular values
$\Sigma^t_{r \times r}$ is a diagonal matrix with singular entries ${\bf \Sigma}_{kk} = 2 (\Omega_H^t)_{kk}^2$ for $k=1,\ldots,r$. The truncation criterion
is $\epsilon \leq (\Omega_H)_{kk}/(\Omega_H)_{11}$ with arbitrary $\epsilon$. The maximum value of $k$ which satisfies the above mentioned criterion gives the rank $r$.

As shown in Eq.~(\ref{Gamma3}), the singular vector ${\bf U_H^t}$ not only gives the singular values of the Fisher matrix but also gives transformation 
into the new phase parameters through linearly combination of $\theta^\alpha$. The new phase parameters are
 \begin{equation}
{\bf \theta' = U^{t \dag}_H \theta} \,.
\end{equation}
where  ${\bf \theta'} \equiv \{\psi'_1,\psi'_2,\psi'_3,\ldots,\psi'_r\}$ is a $r<8$ dimensional phase vector. The Fisher matrix and hence the covariance matrix are diagonal in  ${\bf \theta'}$  space. Further,  the new
phase parameters  ${\bf \theta'}$ are statistically independent variables with zero mean, variance $\sigma_\alpha^2 = (\Sigma^t_{\alpha \alpha})^{-1}$
which provides the absolute error. 
%%%%%%%%%%%%%%%%%%%%%%%%%%%%%%%%%%%%%%%%%%%%%%%%%%%%%%%%%%%%
\subsection{Relative errors for eLISA noise configuration}
%%%%%%%%%%%%%%%%%%%%%%%%%%%%%%%%%%%%%%%%%%%%%%%%%%%%%%%%%%
We apply the above technique for the eLISA  and LISA configurations using the noise curve given in Eqs (1)-(5) of Ref.~\cite{eLISA} and Eqs (2.28)-(2.31) of Ref.~\cite{BBW05a}. We ignore the orbital motion of the space craft and use pattern averaged waveforms. We retain the three most significant new parameters which corresponds to $\epsilon=10^{-6}$.
\begin{figure}[t]
\begin{center}
\includegraphics[scale=0.3]{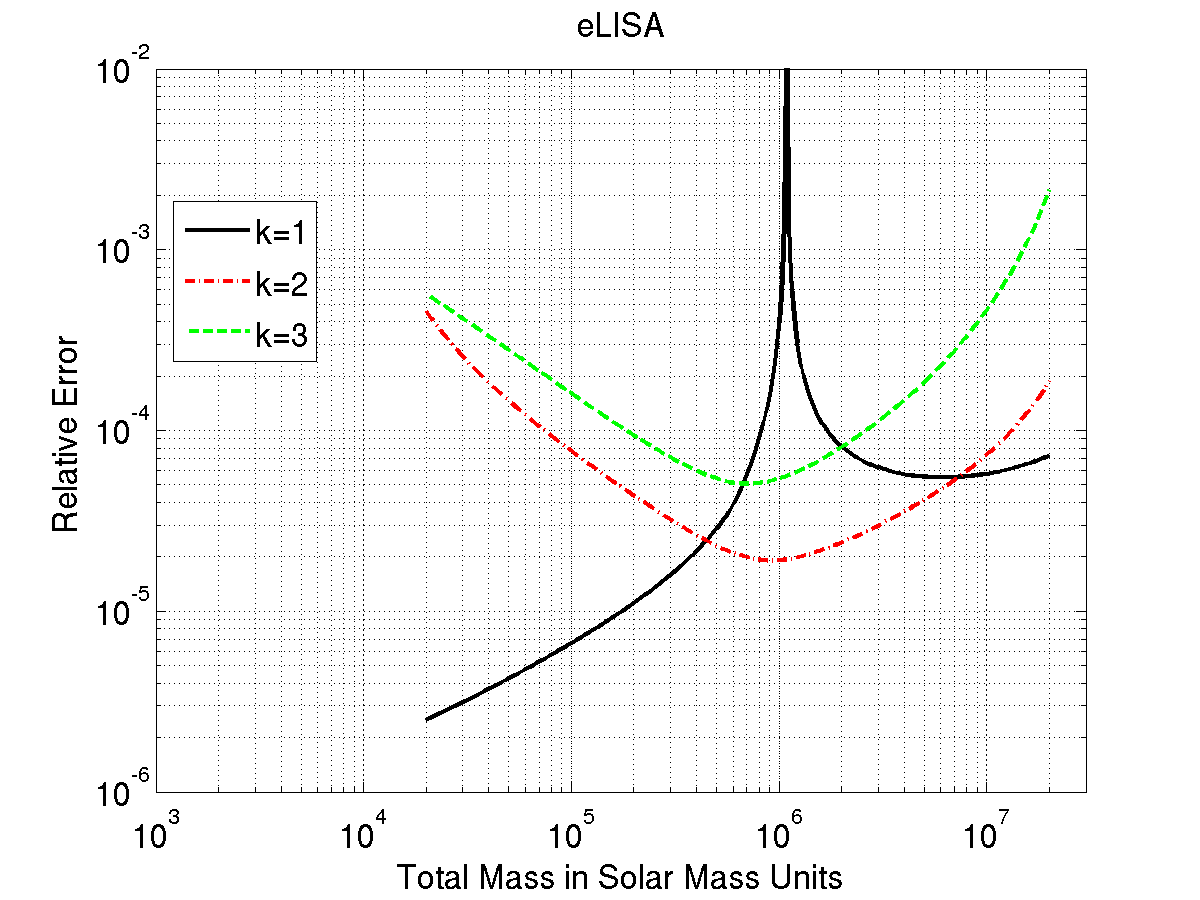}
\includegraphics[scale=0.3]{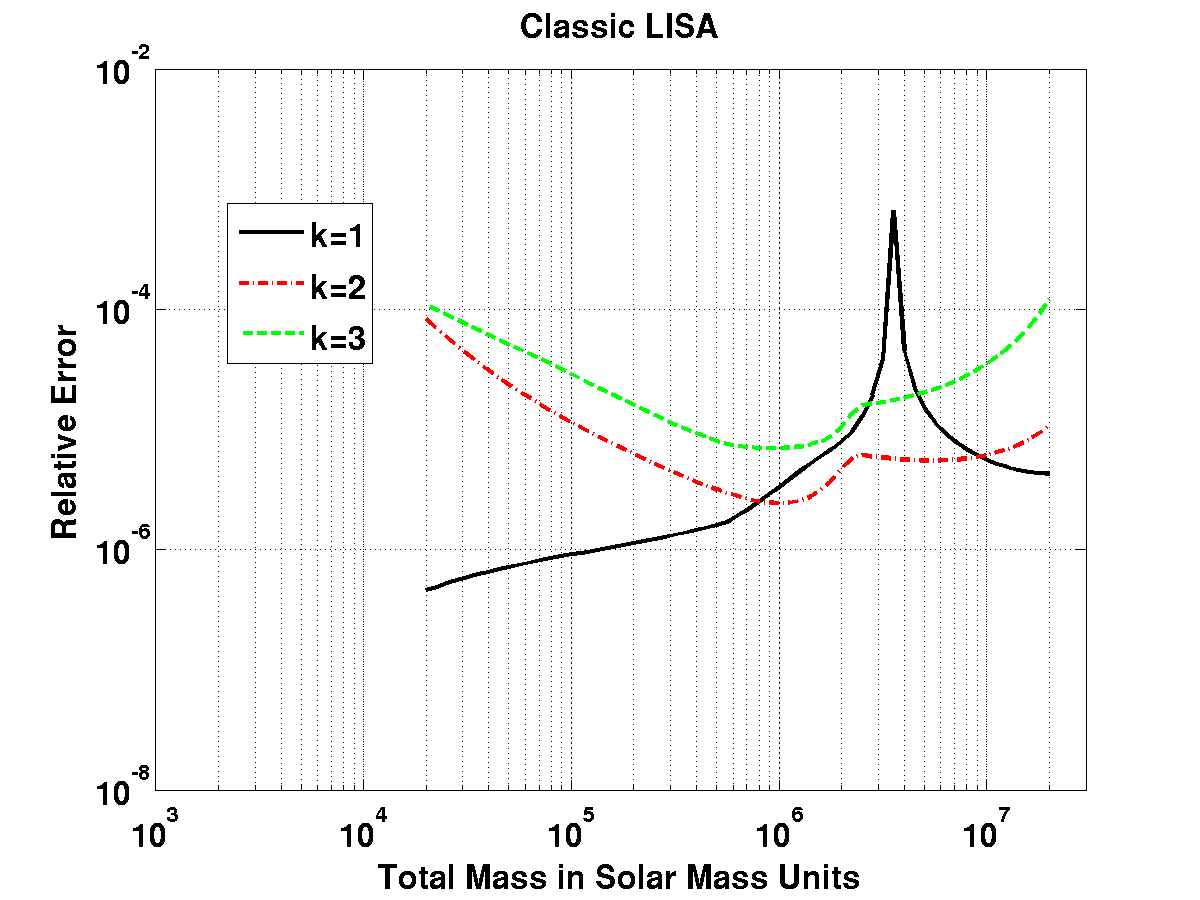}
\caption{Errors in the most dominant new parameters (which are linear combinations of the old  parameters) obtained from SVD in the LISA and eLISA bands as functions of the total mass of the binary for equal mass systems $\eta = 0.25$. Distance to the source is assumed to be 3 Gpc irrespective of the total mass. Spins of the binary systems are ignored.}\label{fig:errors}
\end{center}
\end{figure}

In Fig.~{\ref{fig:errors}}, we plot the relative errors of the new parameters with the total mass for the dominant parameters for eLISA and classic LISA cases. It is obvious that eLISA has poorer estimation of the new parameters roughly an order of magnitude worse than classic LISA for all the three parameters.
This may be attributed to the reduced sensitivity of eLISA, compared to LISA, due to change in the configuration. Except for the spiky segment, all the errors are less than $10^{-3}$ for eLISA and $10^{-4}$ for LISA highlighting the unprecedented accuracies with which these new parameters can be extracted by space-based detectors which are sensitive to the low frequency GWs. Implications of this for tests of GR are currently under investigation. We give below a Table of accuracies with which three leading parameters may be extracted by aLIGO, ET, eLISA and LISA.
\begin{center}
\begin{table}
\tbl{Typical relative errors $\frac{\Delta \Psi_k'}{\Psi_k'}$ of the new phasing coefficients obtained via SVD, for $k=1,2,3$ for aLIGO, ET, eLISA and LISA configurations.}
{\begin{tabular}{cccc}
\hline\hline\\
Detector & $k=1$ & $k=2$ & $k=3$\\
\hline\\
aLIGO & $10^{-4}$ & $10^{-4}$ & $10^{-4}$\\
ET & $10^{-6}$ & $10^{-6}$ & $10^{-6}$\\
eLISA & $10^{-6}$ & $10^{-5}$ & $10^{-4}$\\
LISA & $10^{-6}$ & $10^{-6}$ & $10^{-6}$\\
\hline\hline\\
\end{tabular}}
\end{table}
\end{center}

%Its worth recalling why we needed to invoke the tool of SVD in PTPN. Our basic
%goal is to reduce the parameter space by dropping those parameters which are less dominant. The basic idea is to set a threshold for the principal values below which we will call the parameter is not dominant. Using this criterion we can reduce the dimension of the parameter space restricting only to the most dominant ones. 
%
%%%%%%%%%%%%%%%%%%%%%%%%%%%%%%%%%%%%%%%%%%%%%%%
\section{Conclusion}\label{sec:conclusion}
We have presented an overview of various proposed methods to test GR and alternative theories
of gravity using GW observations of inspiralling compact binaries. We compared,
in each case, the ability of various existing, ongoing and proposed ground-based detectors
and the proposed space-based detectors to constrain theories of gravity. It is very clear that one needs to go the low-frequency GW observations of space-based detectors to test GR with very good accuracy. Hence missions like eLISA/NGO, LISA, DECIGO are extremely important for fundamental physics.

As a final note, we would like to point out that this is {\it not} a comprehensive review of the field. Also there are various very interesting proposals using merger and ringdown waveforms of the compact binaries which is not discussed in this article. 
 \section{Acknowledgements}
KGA would like to thank the organizers of the ASTROD symposium in Bangalore. We thank Bala Iyer, Wei-Tou Ni and Kent Yagi for useful comments on the manuscript. The work is supported in part by AP's MPG-DST Max-Planck India Partner Group Grant.
\bibliographystyle{ws-ijmpd}

\bibliography{ref-list}

\end{document}